\documentclass[aps,pre,reprint,eqsecnum,longbibliography]{revtex4-1}

\usepackage{natbib}

\usepackage{bm}
\usepackage{enumerate}

\usepackage[dvips]{graphicx}
\usepackage{color}
\usepackage{amsmath}
\usepackage{amssymb}
\usepackage{pstricks,pst-grad}
\usepackage{amsbsy}
\usepackage{epic}
\usepackage[normalem]{ulem}

\usepackage[linktoc=all,unicode=true,pdfusetitle,  bookmarks=true,bookmarksnumbered=false,bookmarksopen=false, breaklinks=true,pdfborder={0 0 0},backref=false,colorlinks=true]{hyperref}
\hypersetup{linkcolor=blue,citecolor=red}

\usepackage[hyphenbreaks]{breakurl}

\def\bal#1\eal{\begin{align}#1\end{align}}
\newcommand\beq{\begin{equation}}
\newcommand\eeq{\end{equation}}
\newcommand\beqa{\begin{eqnarray}}
\newcommand\eeqa{\end{eqnarray}}

\newcommand{\sm}{\text{sm}}
\newcommand{\nn}{\nonumber\\}

\newcommand{\thr}{\text{th}}

\newcommand{\al}{\alpha}

\newcommand{\bt}{\widetilde\beta}
\newcommand{\at}{\widetilde\alpha}

\newcommand{\taus}{s}
\newcommand{\so}{\omega}
\newcommand{\Th}{\Theta}
\newcommand{\mitr}{\text{M}}

\begin{document}
\title{Impact of roughness on the instability of a free-cooling granular gas}

\author{Vicente Garz\'o}
\email{vicenteg@unex.es} \homepage{http://www.eweb.unex.es/eweb/fisteor/vicente/}
\author{Andr\'es Santos}
\email{andres@unex.es} \homepage{http://www.eweb.unex.es/eweb/fisteor/andres/}
\affiliation{Departamento de F\'{\i}sica and Instituto de Computaci\'on Cient\'{\i}fica Avanzada (ICCAEx), Universidad de Extremadura, E-06006 Badajoz, Spain}
\author{Gilberto M. Kremer}
\email{kremer@fisica.ufpr.br}
\affiliation{Departamento de F\'{\i}sica, Universidade Federal do Paran\'a, 81531-980 Curitiba, Brazil}

\begin{abstract}
A linear stability analysis of the hydrodynamic equations with
respect to the homogeneous cooling state is carried out to identify
the conditions for stability of a granular gas of rough hard spheres. The description is based on the results for the transport coefficients  derived from the Boltzmann equation for inelastic rough hard spheres [Phys.\ Rev.\ E \textbf{90}, 022205 (2014)], which take into account the complete nonlinear dependence of the transport coefficients and the cooling rate on the coefficients of normal and tangential restitution. As expected, linear
stability analysis shows that a doubly degenerate transversal (shear) mode and a longitudinal (``heat'') mode are unstable with respect to long enough wavelength excitations. The instability is driven by the shear mode above a certain inelasticity threshold; at larger inelasticity, however, the instability is driven by the heat mode for an inelasticity-dependent range of medium roughness. Comparison with the case of a granular gas of inelastic smooth spheres confirms previous simulation results about the dual role played by surface friction: while small and large levels of roughness make the system less unstable than the frictionless system, the opposite happens at medium roughness. On the other hand, such an intermediate window of roughness values shrinks as inelasticity increases and eventually disappears at a certain value,  beyond which the rough-sphere gas is always less unstable than the smooth-sphere gas. A comparison with some preliminary simulation results shows a very good agreement for conditions of practical interest.

\end{abstract}

\date{\today}
\maketitle

\section{Introduction}
\label{sec1}

One of the most characteristic features of granular fluids, as compared with ordinary fluids, is the spontaneous formation of velocity vortices and density clusters in freely evolving flows [homogenous cooling state (HCS)]. This kind of flow instability  originates by the dissipative nature of particle-particle collisions \cite{FH17}. In the case of smooth and frictionless inelastic hard spheres, those instabilities were first detected independently, by means of computer simulations,  by Goldhirsch and Zanetti \cite{GZ93} and McNamara \cite{M93b}. An important property of the clustering instability is that it is restricted  to long-wavelength excitations, and, hence, it can be suppressed for systems that are small enough. In addition, simulations also show that vortices generally preempt  clusters, so that the onset of instability is generally associated with the transversal shear modes.

Instabilities of freely cooling flows of smooth spheres can be well described by means of a linear stability analysis of the Navier--Stokes (NS) hydrodynamic equations. This analysis gives a critical wave number $k_c$ and an associated critical wavelength $L_c=2\pi/k_c$, such that the system becomes unstable if its linear size is larger than $L_c$. The evaluation of $L_c$ requires knowledge of the dependence of the NS transport coefficients on the coefficient of normal restitution $\al$, and therefore the determination of $L_c$ is perhaps one of the most interesting applications of the NS equations. In addition,  comparison between kinetic theory and computer simulations for the critical size can be considered  a clean way of assessing the degree of accuracy of the former, since theoretical results are obtained in most of the cases under certain approximations (e.g., by considering the leading terms in a Sonine polynomial expansion).

The accuracy of the kinetic-theory prediction of $L_c$  for a low-density monodisperse granular gas of inelastic smooth hard spheres \cite{BDKS98,G05} has been verified by means of the direct simulation Monte Carlo DSMC method \cite{BRM98} and, more recently, by molecular dynamics (MD) simulations for a granular fluid at moderate density \cite{MDCPH11,MGHEH12}. Similar comparisons have also been made for polydisperse systems \cite{GD02,GMD06,GDH07,GHD07} in the low-density regime \cite{BR13} and moderate densities \cite{MGH14}. In general, the theoretical predictions for the critical size compare quite well with computer simulations, even for strong dissipation.

On the other hand, all the above works ignore the influence of collisional friction on instability. Needless to say, grains in nature are typically frictional, and, hence, energy transfer between the translational and rotational degrees of freedom occurs upon particle collisions. The simplest way of accounting for friction is, perhaps, through a constant coefficient of tangential restitution $\beta$, ranging from $-1$ (perfectly smooth spheres) to $1$ (perfectly rough spheres).
To the best of our knowledge, the first work where the impact of roughness on a transport coefficient (self-diffusion) was investigated for
arbitrary inelasticity and roughness was carried out by Bodrova and Brilliantov \cite{BB12}.
On the other hand, in contrast to the case of smooth-sphere granular gases \cite{BDKS98,GD99,L05,GD02,GDH07,GHD07,GST18}, most of the attempts made for evaluating the other transport coefficients of inelastic rough hard spheres have been restricted to nearly elastic collisions ($\al \lesssim 1$) and either nearly smooth particles ($\beta \gtrsim -1$) \cite{JR85a,GNB05,GNB05b} or nearly perfectly rough particles ($\beta \lesssim 1$) \cite{JR85a,L91}. An extension of the previous works to arbitrary values of $\al$ and $\beta$ has been recently carried out by Kremer \emph{{et al.}} \cite{KSG14} for a \emph{dilute} granular gas. Explicit expressions for the NS transport coefficients and the cooling rate were obtained as nonlinear functions of both coefficients of restitution and the moment of inertia. The knowledge of the transport coefficients opens up the possibility of performing a linear stability analysis of the hydrodynamic equations with respect to the HCS state to identify the conditions for stability as functions of the wave vector and the coefficients of restitution.

A previous interesting work on flow instabilities for a dense granular gas of inelastic and rough hard spheres was carried out by Mitrano {\emph{et al.}}  \cite{MDHEH13}. In that paper, the authors compare their MD simulations against theoretical predictions obtained from a simple stability analysis where friction is only accounted for through its impact on the cooling rate, since the NS transport coefficients are otherwise assumed to be formally the same as those obtained for frictionless particles \cite{GD99}. In spite of the simplicity of this theoretical approach, the obtained predictions compare well with MD simulations. In particular, they observe that, paradoxically, high levels of roughness can actually attenuate instabilities relative to the frictionless case.

Even though, as said above,  the predictions made by Mitrano {\emph{et al.}}  \cite{MDHEH13} agree reasonably well with simulation results, they lack a sounder basis as they are essentially based on the hydrodynamic description of smooth particles and only the cooling rate incorporates the complete nonlinear dependence on $\beta$. As a matter of fact, the derivation of the transport coefficients for rough spheres \cite{KSG14} was not known before Ref.\ \cite{MDHEH13} was published. Therefore, it is worthwhile  assessing  to what extent the previous results are indicative to what happens when the complete nonlinear dependence of both $\al$ and $\beta$ on the transport coefficients is included in the stability analysis. The main aim of this paper is to fill this gap by revisiting the problem of the HCS stability of a granular gas described by the Boltzmann kinetic equation, this time making use of the knowledge of the NS transport coefficients of a granular gas of inelastic rough hard spheres \cite{KSG14}.

As expected, linear stability analysis shows two transversal (shear) modes and a longitudinal (``heat'') mode to be unstable with respect to long enough wavelength excitations. The corresponding critical values for the shear and heat modes are explicitly determined as functions of the (reduced) moment of inertia and the coefficients of restitution $\al$ and $\beta$. The results show that the instability is mainly driven by the transversal shear mode, except for values of the coefficient of normal restitution $\alpha$ smaller than a certain threshold value $\alpha_\thr$. Below that value, the instability is dominated by the heat mode, provided that the coefficient of tangential restitution $\beta$ lies inside an $\alpha$-dependent window of values around $\beta\approx 0$. Moreover, below an even smaller value $\alpha_D$, any value of $\beta$ is enough to attenuate the instability effect with respect to the frictionless case.

The plan of the paper is as follows. The NS hydrodynamic equations for a dilute granular gas of inelastic and rough hard spheres are shown in Sec.\ \ref{sec2}, the explicit expressions for the transport coefficients being given in the Appendix. Next, the linear stability analysis of the NS equations is worked out in Sec.\ \ref{sec3}, where it is found that the two transversal shear modes are decoupled from the three longitudinal ones.
Section \ref{sec4} deals with a detailed discussion of the results, with a special emphasis on the impact of roughness on the instability of the HCS.
The paper ends in Sec.\ \ref{sec5} with a summary and concluding remarks.

\section{Navier--Stokes hydrodynamic equations}
\label{sec2}

We consider a dilute granular gas composed of inelastic and rough hard
spheres of mass $m$, diameter $\sigma$, and moment of inertia $I=\kappa m\sigma^2/4$ (where the dimensionless parameter $\kappa$ characterizes the mass distribution in a sphere).  Since, in general,  particle rotations and surface friction are relevant in the description of granular flows, collisions are characterized by constant coefficients of normal restitution ($\al$) and tangential restitution ($\beta$).
The coefficient $\al$ measures the postcollisional shrinking in the magnitude of the normal component of the relative velocity of the two points at contact; the coefficient $\beta$ does the same but for the tangential component \cite{SKG10,VLSG17b}.
As said in Sec.\ \ref{sec1}, the coefficient $\al$ ranges from $0$ (perfectly inelastic particles) to $1$ (perfectly elastic particles) while the coefficient $\beta$  ranges from $-1$ (perfectly smooth spheres) to $1$ (perfectly rough spheres).

At a kinetic level, all the relevant information
on the system is given through the one-particle velocity distribution function, which is assumed to obey the (inelastic) Boltzmann equation \cite{BP04,RN08}. From it, one can derive the (macroscopic) hydrodynamic balance equations
for the particle number density $n({\mathbf r}, t)$,  the local temperature $T({\mathbf r}, t)$, and the flow velocity ${\mathbf u}({\bf r}, t)$:
\begin{subequations}
\label{2.1-2.3}
\begin{equation}
D_{t}n+n \nabla \cdot \mathbf{u}=0,  \label{2.1}
\end{equation}
\begin{equation}
D_{t}T+\frac{1}{3n}\left(\nabla \cdot \mathbf{q}+\mathsf{P}:\nabla \mathbf{u}\right)+\zeta T=0,
\label{2.3}
\end{equation}
\begin{equation}
mn D_{t}\mathbf{u}+\nabla \cdot \mathsf{P}=0.
\label{2.2}
\end{equation}
\end{subequations}
In Eqs.\ \eqref{2.1-2.3}, $D_{t}=\partial _{t}+\mathbf{u}\cdot \nabla $
is the material derivative,   ${\sf P}$ is the pressure tensor, ${\bf q}$ is the heat flux, and $\zeta$ is the cooling rate due to the collisional energy dissipation.

The balance equations \eqref{2.1-2.3} do not constitute a closed set of equations for the hydrodynamic fields $\{n,T,\mathbf{u}\}$, unless the fluxes and the cooling rate are further specified in terms of the hydrodynamic fields and their gradients. As mentioned in Sec.\ \ref{sec1}, the detailed form of the constitutive equations and the transport coefficients appearing in them have recently been obtained by applying the Chapman--Enskog method to the Boltzmann equation. To first order in the gradients, the corresponding constitutive equations are \cite{KSG14}
\begin{subequations}
\label{2.4-2.6}
\begin{equation}
P_{ij}=nT \tau_t\delta_{ij}-\eta \left( \nabla _{j}u_{i }+\nabla_{i}u_{j}-\frac{2}{3}\delta_{ij}
\nabla \cdot \mathbf{u}\right)-\eta_b \delta_{ij} \nabla\cdot \mathbf{u},
 \label{2.4}
\end{equation}
\begin{equation}
\mathbf{q}=-\lambda \nabla T-\mu \nabla n,
\label{2.5}
\end{equation}
\begin{equation}
\label{2.6}
\zeta=\zeta^{(0)}-\xi \nabla \cdot \mathbf{u}.
\end{equation}
\end{subequations}
Here, $\tau_t\equiv T_t^{(0)}/T$ is the ratio between the (partial) translational temperature $T_t^{(0)}$ and the granular temperature $T$ in the reference HCS, $\eta$ is the shear viscosity, $\eta_b$ is the bulk viscosity, $\lambda$ is the thermal
conductivity,  $\mu$ is a new transport coefficient (heat diffusivity coefficient or Dufour-like coefficient), not present
in the conservative case, and $\xi$ characterizes the NS correction to the HCS cooling rate $\zeta^{(0)}$. The last three coefficients have both translational ($\lambda_t$, $\mu_t$, $\xi_t$) and rotational ($\lambda_r$, $\mu_r$, $\xi_r$) contributions.
Moreover, the (partial) rotational temperature $T_r^{(0)}$ in the HCS is given by $T_r^{(0)}=\tau_r T$, where $\tau_r=2-\tau_t$.

Insertion of the NS constitutive equations \eqref{2.4-2.6}
into the balance equations \eqref{2.1-2.3} yields the corresponding NS
hydrodynamic equations for $n$,  $T$, and ${\bf u}$:
\begin{subequations}
\label{2.15-2.17}
\begin{equation}
\label{2.15.1}
D_tn=-n\nabla \cdot \mathbf{u},
\end{equation}
\begin{eqnarray}
\label{2.17}
\left(D_t+\zeta^{(0)}\right)T&=&\left(\xi-\frac{\tau_t}{3}\right)T\nabla \cdot {\bf u}+\frac{1}{3n}\nabla \cdot \left
(\lambda \nabla T+\mu \nabla n\right)\nonumber\\
& & +\frac{1}{3n}\Big[\eta \left(\nabla_iu_j+\nabla_ju_i\right)-\left(\frac{2\eta}{3}-\eta_b\right)
\nonumber\\
& & \times \delta_{ij}\nabla \cdot \mathbf{u}
\Big]\nabla_i u_j,
\end{eqnarray}
\beqa
\label{2.16}
mn D_t u_i &=&-\tau_t \nabla_i(n T)+\nabla_j\Big[\eta \left(\nabla_iu_j+\nabla_ju_i \right)
\nonumber\\
& &
-\left(\frac{2\eta}{3}-\eta_b\right)
\delta_{ij}\nabla \cdot \mathbf{u}\Big].
\eeqa
\end{subequations}

As noted in previous stability studies \cite{G05,GMD06}, in principle one should consider  terms up to second order in the gradients in Eq.\ \eqref{2.6} for the cooling rate, since this is the order of the terms of Eq.\ \eqref{2.17} coming from the pressure tensor and the heat flux. On the other hand, it has been shown for  dilute granular gases of smooth spheres that the second order contributions to the cooling rate are negligible as compared with  their corresponding zeroth- and first-order contributions \cite{BDKS98}. It is assumed here that the same happens for inelastic rough hard spheres.

To close the hydrodynamic problem posed by Eqs.\ \eqref{2.15-2.17}, one needs to express the coefficients, $\tau_t$, $\zeta^{(0)}$, $\eta$, $\eta_b$, $\lambda$, $\mu$, and $\xi$ as functions of the hydrodynamic fields ($n$ and $T$) and of the mechanical parameters ($\alpha$,  $\beta$, and $\kappa$). The former dependence is dictated by dimensional analysis, while the dependence on $\alpha$, $\beta$, and $\kappa$ requires resorting to approximations.

In the two-temperature Maxwellian approximation, the HCS parameters $\tau_t$, $\tau_r$, and $\zeta^{(0)}$ are given by \cite{GS95,LHMZ98,SKG10,VLSG17b}
\begin{subequations}
\label{2.11-2.10}
\beq
\label{2.11}
\tau_t=\frac{2}{1+\theta}, \quad \tau_r=\frac{2\theta}{1+\theta}, \quad \theta=\sqrt{1+h^2}+h,
\eeq
\beq
\label{2.10}
\zeta^{(0)}=\zeta^*\nu,\quad \nu=\frac{16}{5}n\sigma^2\chi(\phi)\sqrt{\frac{\pi T \tau_t}{m}},
\eeq
\end{subequations}
where  $\theta=T_r^{(0)}/T_t^{(0)}$ is the rotational-to-translational temperature ratio and
\begin{subequations}
\label{2.12-2.13}
\beq
\label{2.12}
h=\frac{(1+\kappa)^2}{2\kappa(1+\beta)^2}
\left[1-\al^2-(1-\beta^2)\frac{1-\kappa}{1+\kappa}\right],
\eeq
\beq
\label{2.13}
\zeta^*=\frac{5}{12}\frac{1}{1+\theta}\left[1-\al^2+(1-\beta^2)\frac{\kappa+\theta}{1+\kappa}\right].
\eeq
\end{subequations}
In the second equality of Eq.\ \eqref{2.10}, $\chi(\phi)$ is the contact value of the pair correlation function (Enskog factor), which is a function of the solid volume fraction $\phi\equiv\frac{\pi}{6}n\sigma^3$. A simple and accurate prescription is $\chi(\phi)=(1-\phi/2)/(1-\phi)^3$ \cite{CS69}. In principle, the results of this paper apply to the Boltzmann limit $\phi\to 0$, in which case $\chi(\phi)\to 1$. Nevertheless, keeping the Enskog factor $\chi(\phi)$ in the collision frequency $\nu$ allows us to account for basic excluded volume effects which are present if $\phi$ is small but not strictly zero.

From Eqs.\ \eqref{2.11} and \eqref{2.12}, we note that  $\theta$ diverges in the quasismooth limit $\beta\to -1$ as
\beq
\label{theta_quasi}
\theta =\frac{(1-\alpha^2)(1+\kappa)^2}{\kappa}(1+\beta)^{-2}.
\eeq
Consequently, in that limit the reduced cooling rate vanishes as
\beq
\label{zeta_quasi}
\zeta^*= \frac{5}{6(1+\kappa)}(1+\beta).
\eeq
This contrasts with the \emph{purely smooth}  case \cite{BDKS98}, where
\beq
\label{zeta_smooth}
\zeta^*_\sm=\frac{5}{12}(1-\alpha^2),
\eeq
which can be obtained from Eq.\ \eqref{2.13} by setting $\beta=-1$ and formally taking $\theta=0$.
The difference between Eqs.\ \eqref{zeta_quasi} and \eqref{zeta_smooth} illustrates the strong singular character of the limit $\beta\to -1$ in the HCS \cite{BPKZ07,KBPZ09,SKG10,SKS11,S11b}.

The expressions of the NS transport coefficients in the first Sonine approximation are \cite{KSG14}
\beq
\label{2.7}
\eta=\eta_0 \eta^*,\quad \eta_b=\eta_0 \eta_b^*,
\quad \lambda=\lambda_0 \lambda^*, \quad \mu=\frac{T\lambda_0}{n} \mu^*,
\eeq
where
\beq
\label{2.9}
\eta_0=\frac{n T \tau_t}{\nu}, \quad \lambda_0=\frac{15}{4}\frac{\eta_0}{m}
\eeq
are the shear viscosity and the thermal conductivity coefficients, respectively, of a gas of elastic and smooth spheres at the (translational) temperature $T\tau_t$.
The dimensionless quantities $\eta^*$, $\eta_b^*$, $\lambda^*$, and $\mu^*$ are nonlinear
functions of $\kappa$, $\alpha$, and $\beta$ given by
\begin{subequations}
\label{2.14-2.15}
\beq
\label{2.14}
\eta^*=\frac{1}{\nu_\eta^*-\frac{1}{2}\zeta^*}, \quad \eta_b^*=\tau_r \gamma_E,
\eeq
\beq
\label{2.15}
\lambda^*=\frac{2}{3}\tau_t\gamma_{A_t}+\frac{2}{5}\tau_r\gamma_{A_r}, \quad
\mu^*=\frac{2}{3}\tau_t\gamma_{B_t}+\frac{2}{5}\tau_r\gamma_{B_r},
\eeq
\end{subequations}
where  the explicit forms of the quantities $\nu_\eta^*$, $\gamma_E$, $\gamma_{A_t}$, $\gamma_{A_r}$, $\gamma_{B_t}$, and $\gamma_{B_r}$ are displayed in the Appendix [see Eqs.\ \eqref{a1-a2} and \eqref{a6-a9}].

Finally, the NS cooling-rate coefficient $\xi$ is
\beq
\label{2.13.1}
\xi=\gamma_E \Xi,
\eeq
where  $\Xi$ is given by Eq.\ \eqref{a14}.

In the purely smooth case ($\beta=-1$), one has $\tau_t=2$, $\tau_r=0$, $\xi=0$, $\eta_b^*=0$, $\zeta^*=\zeta^*_\sm$, $\eta^*=\eta^*_\sm$, $\lambda^*=2\lambda_\sm^*$, and $\mu^*=2\mu_\sm^*$, with
\begin{subequations}
\beq
\eta^*_\sm=\frac{24}{(1+\alpha)(13-\alpha)},
\eeq
\beq
\lambda_\sm^*=\frac{32}{(1+\alpha)(9+7\alpha)},
\eeq
\beq
\mu_\sm^*=\frac{640(1-\alpha)}{(1+\alpha)(9+7\alpha)(19-3\alpha)}.
\eeq
\end{subequations}

\section{Linear stability analysis of the NS equations}
\label{sec3}

It is well known that the NS equations admit a simple solution corresponding to the so-called HCS. It describes a uniform state with vanishing flow field and a temperature $T$ decreasing monotonically in time,
\begin{equation}
\label{3.1}
\nabla n_H=\nabla T_H=0, \quad \partial_t
\ln T_H=-\zeta^{(0)}_{H}, \quad {\bf u}_H={\bf 0},
\end{equation}
where henceforth the subscript $H$ denotes quantities evaluated in the HCS.

On the other hand, the HCS is known to become unstable under sufficiently long wavelength excitations \cite{GZ93,M93b,BDKS98,G05,BRM98,MDCPH11,MGHEH12,GD02,GMD06,GDH07,GHD07,BR13,MGH14,MDHEH13}. Our aim here is to study the stability conditions of the HCS by means of a linear stability analysis of Eqs.\ \eqref{2.15-2.17}  for small initial excitations with respect to the homogeneous state. Thus, we assume that the deviations $\delta y_{a}(\mathbf{r},t)=y_{a}(\mathbf{r},t)-y_{H a}(t)$ of the hydrodynamic fields $\{y_a;a=1,\ldots,5\}\equiv \{n,T,\mathbf{u}\}$ from their values in the HCS are small, and neglect terms of second and higher order. The resulting equations  for $\delta n$, $\delta T$, and $\delta u_i$  are
\begin{subequations}
\label{3.3-3.5}
\beq
\label{3.3}
\partial_t \frac{\delta n}{n_H}=-v_H\nabla \cdot  \frac{\delta\mathbf{u}}{v_H},
\eeq
\bal
\label{3.5}
\partial_t\frac{\delta T}{T_H}=&-\zeta_H^{(0)}\left(\frac{\delta n}{n_H}+\frac{1}{2}\frac{\delta T}{T_H}\right)+\left(\xi-\frac{\tau_t}{3} \right) v_H\nabla \cdot  \frac{\delta\mathbf{u}}{v_H}\nn
&+\frac{\lambda_{0H}}{3n_H}\nabla^2\left(\lambda^* \frac{\delta T}{T_H}
+\mu^*\frac{\delta n}{n_H}\right),
\eal
\bal
\label{3.4}
\partial_t \frac{\delta u_i}{v_H}=&\frac{\zeta_{H}^{(0)}}{2} \frac{\delta u_i}{v_H}-v_H\nabla_i\left( \frac{\delta n}{n_H}+
\frac{\delta T}{T_H}\right)\nn
&+\frac{\eta_{0H}}{mn_H}\left[\left(\frac{\eta^*}{3}+\eta_b^*\right)\nabla_i \nabla \cdot \frac{\delta \mathbf{u}}{v_H}
+\eta^* \nabla^2 \frac{\delta u_i}{v_H}\right],
\eal
\end{subequations}
Note that in Eqs.\ \eqref{3.3-3.5} the deviations $\{\delta n,\delta T, \delta \mathbf{u}\}$ have been scaled with respect to the HCS quantities $\{n_H,T_H(t),v_H(t)\}$, where
\beq
\label{v_H}
v_H(t)=\sqrt{\frac{T_{H}(t)\tau_t}{m}}
\eeq
is the thermal velocity in the HCS. This scaling is nontrivial in the cases of the flow velocity and the temperature since the reference state is cooling down and thus $v_H$ and $T_H$ are time-dependent quantities.

As in the purely smooth case ($\beta=-1$) \cite{BDKS98,G05}, it is advisable to introduce the scaled variables
\begin{equation}
\label{3.2}
\taus(t)=\frac{1}{2}\int_0^t dt'\,\nu_{H}(t'), \quad {\boldsymbol{\ell}}=\frac{1}{2}
\frac{\nu_{H}(t)}{v_{H}(t)}\mathbf{r},
\end{equation}
where $\nu_H(t)$ is defined by Eq.\ \eqref{2.10} with $T\to T_H$ evaluated in the HCS. The quantity $\taus(t)$ is a measure of the number of collisions per particle up to time $t$, while $\boldsymbol{\ell}$ measures position in units of the mean free path (notice that the ratio $\nu_H/v_H$ is independent of time).
A set of Fourier transformed dimensionless variables are then
defined by
\begin{equation}
\label{3.6}
\rho_{{\bf k}}(\taus)=\frac{\delta \widetilde{n}_{{\bf k}}(\taus)}{n_{H}}, \quad \Th_{{\bf k}}(\taus)=\frac{\delta
\widetilde{T}_{{\bf k}}(\taus)}{T_{H}(\taus)},\quad
{\bf w}_{{\bf k}}(\taus)=\frac{\delta \widetilde{{\bf u}}_{{\bf
k}}(\taus)}{v_H(\taus)},
\end{equation}
where the Fourier transforms $\{\delta \widetilde{y}_{{\bf k},a}(s);a=1,\ldots,5\}\equiv \{\delta \widetilde{n}_{{\bf k}}(s), \delta \widetilde{T}_{{\bf k}}(\taus),\delta\widetilde{\bf
u}_{\bf k}(\taus)\}$ are
\begin{equation}
\label{3.7}
\delta \widetilde{y}_{{\bf k},a}(\taus)=\int d{\boldsymbol {\ell}}\,
e^{-\imath{\bf k}\cdot {\boldsymbol {\ell}}}\delta y_{a}
({\boldsymbol {\ell}},\taus).
\end{equation}
Note that in Eq.\ \eqref{3.7} the wave vector $\mathbf{k}$ is dimensionless.

In terms of the variables \eqref{3.2} and \eqref{3.6}, Eqs.\ \eqref{3.3-3.5} become
\begin{subequations}
\beq
\partial_\taus \rho_{\mathbf{k}}=-\imath \mathbf{k}\cdot \mathbf{w}_{\mathbf{k}},
\eeq
\bal
\partial_\taus\Theta_{\mathbf{k}}=&-\zeta^*\left(2\rho_{\mathbf{k}}+\Theta_{\mathbf{k}}\right)+\left(\xi-\frac{\tau_t}{3} \right)\imath\mathbf{k} \cdot  \mathbf{w}_{\mathbf{k}}\nn
&-\frac{5}{8}k^2\left(\lambda^* \Theta_{\mathbf{k}}
+\mu^*\rho_{\mathbf{k}}\right),
\eal
\bal
\partial_\taus w_{\mathbf{k},i}=&\zeta^* w_{\mathbf{k},i}-\imath k_i\left( \rho_{\mathbf{k}}+\Theta_{\mathbf{k}}\right)\nn
&-\frac{1}{2}\left[\left(\frac{\eta^*}{3}+\eta_b^*\right)k_i \mathbf{k} \cdot  \mathbf{w}_{\mathbf{k}}
+\eta^* k^2 w_{\mathbf{k},i}\right].
\eal
\end{subequations}
As expected, the two transversal velocity components
${\bf w}_{{\bf k},\perp}={\bf w}_{{\bf k}}-({\bf w}_{{\bf k}}\cdot
\widehat{{\bf k}})\widehat{{\bf k}}$ (orthogonal to the wave vector ${\bf k}$)
decouple from the other three modes and hence can be analyzed independently. Their evolution equation is
\begin{equation}
\label{3.8}
\left(\partial_\taus-\zeta^*+\frac{\eta^*}{2}
k^2\right){\bf w}_{{\bf k},\perp}=0,
\end{equation}
whose solution is
\begin{equation}
\label{3.9}
{\bf w}_{{\bf k},\perp}(\taus)={\bf w}_{{\bf k},\perp}(0)e^{\so_{\perp}(k)\taus},
\end{equation}
where
\begin{equation}
\label{3.10}
\so_{\perp}(k)=\zeta^*-\frac{\eta^*}{2} k^2.
\end{equation}
This identifies two shear (transversal) modes analogous to the ones for molecular gases. According to Eq.\ \eqref{3.10}, there exists
a critical wave number $k_{\perp}$, given by
\begin{equation}
\label{3.11}
k_{\perp}=\sqrt{\frac{2\zeta^*}{\eta^*}},
\end{equation}
separating  two regimes: shear modes with $k> k_{\perp}$ always decay,
while those with $k< k_{\perp}$ grow exponentially.
For purely smooth spheres, the critical wave number $k_{\perp,\sm}$ is given by Eq.\ \eqref{3.11} with $\zeta^*\to\zeta^*_\sm$ and $\eta^*\to\eta^*_\sm$.

The remaining (longitudinal) modes correspond to $\rho_{{\bf k}}$, $\Th_{{\bf k}}$, and
the longitudinal velocity component of the velocity field, $w_{{\bf k},||}={\bf w}_{{\bf
k}}\cdot \widehat{{\bf k}}$ (parallel to ${\bf k}$). These modes are coupled and obey the matrix equation
\begin{equation}
\partial_\taus \delta y_{{\bf k},a}(\taus )=M_{ab}(k)
 \delta y_{{\bf k},b}(\taus),
\label{3.12}
\end{equation}
where $\delta y_{{\bf k},a}(\taus)$ denotes now the set  $\left\{\rho_{{\bf k}},\Th_{{\bf k}},
 w_{{\bf k},||}\right\}$ and $\mathsf{M}(k)$ is the square matrix
\begin{equation}
\mathsf{M}(k)=-\left(
\begin{array}{ccc}
0 & 0 & \imath k \\
2\zeta^*+\frac{5\mu^*}{8}k^2&\zeta^*+\frac{5\lambda^*}{8}k^2 & \frac{\tau_t-3\xi}{3}\imath k\\
\imath k & \imath k &\frac{4\eta^*+3\eta_b^*}{6}k^2-\zeta^{*}
\end{array}
\right).
  \label{3.13}
\end{equation}

The longitudinal three modes have the form $\exp[\so_{\|,a}(k) \taus]$ for $a=1,2,3$, where
$\{\so_{\|,a}(k)\}$ are the eigenvalues of the matrix ${\sf M}(k)$, i.e.,
they are the solutions of the cubic equation
\beq
\label{3.14}
\so^3+A(k)\so^2+B(k)\so+C(k)=0,
\eeq
where
\begin{subequations}
\label{3.15-17}
\beq
\label{3.15}
A(k)=\left(\frac{5\lambda^*}{8}+\frac{2\eta^*}{3}+\frac{\eta_b^*}{2}\right)k^2,
\eeq
\bal
\label{3.16}
B(k)=&\frac{5\lambda^* }{8}\left(\frac{2\eta^*}{3}+\frac{\eta_b^*}{2}\right)k^4+\Big[1-\xi+\frac{\tau_t}{3}
\nonumber\\
&
+\zeta^*\left(\frac{2\eta^*}{3}+\frac{\eta_b^*}{2}-\frac{5\lambda^*}{8}\right)\Big]k^2-\zeta^{*2},
\eal
\beq
\label{3.17}
C(k)=\frac{5}{8}(\lambda^*-\mu^*)k^4-\zeta^*k^2.
\eeq
\end{subequations}

In the purely smooth case ($\beta=-1$),  Eqs.\ \eqref{3.8}--\eqref{3.15-17} agree with previous results with $\zeta\to\zeta^{(0)}$ \cite{BDKS98}.

\begin{figure*}[tbp]
\includegraphics[width=1.8\columnwidth]{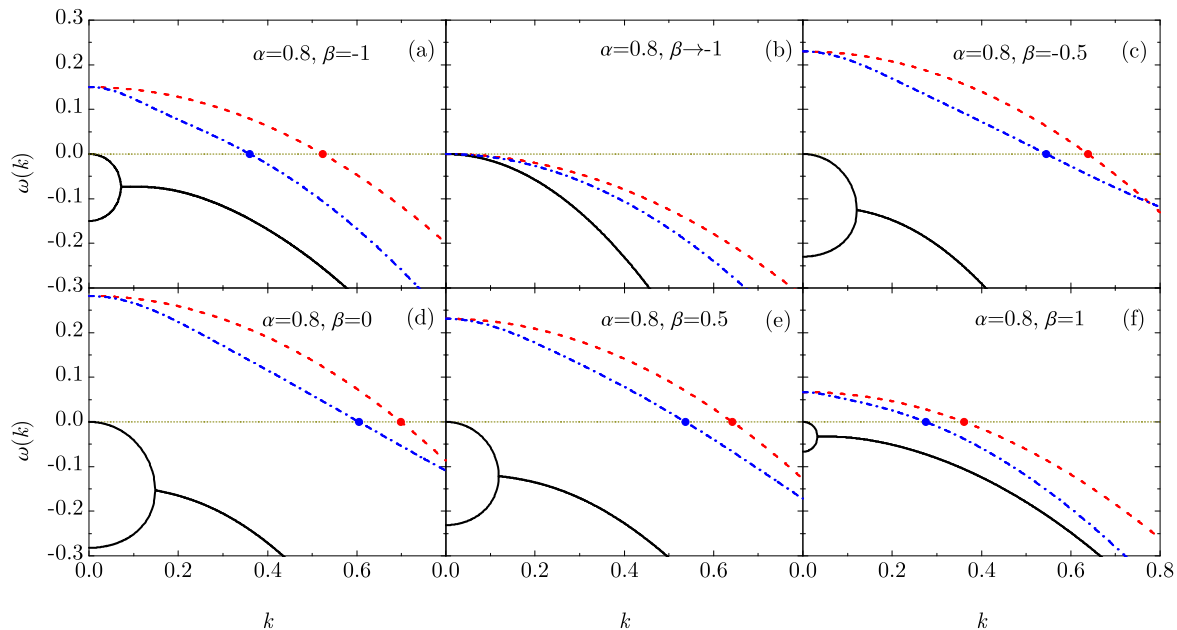}
\caption{Dispersion relations $\omega(k)$ for the hydrodynamic modes vs the
reduced wave number $k$. The curves correspond
to the degenerate shear mode $\omega_\perp$ (-- -- --), the heat mode $\omega_{\|,3}$ (--$\cdot$--$\cdot$--), and the sound modes  $\omega_{\|,1}$ and $\omega_{\|,2}$ (---). Note that when $\omega_{\|,1}$ and $\omega_{\|,2}$ become a complex conjugate pair, only the (common) real part is plotted. The circles denote the critical wave numbers $k_\perp$ and $k_{\|}$. The coefficient of normal restitution is $\alpha=0.8$ and the reduced moment of inertia is $\kappa=\frac{2}{5}$, while the coefficients of tangential restitution are (a) $\beta=-1$ (purely smooth spheres), (b) $\beta\to -1$ (quasismooth limit), (c) $\beta=-0.5$, (d) $\beta=0$, (e) $\beta=0.5$, and (f) $\beta=1$.}
\label{fig:modes}
\end{figure*}

\begin{figure}[tbp]
\includegraphics[width=0.9\columnwidth]{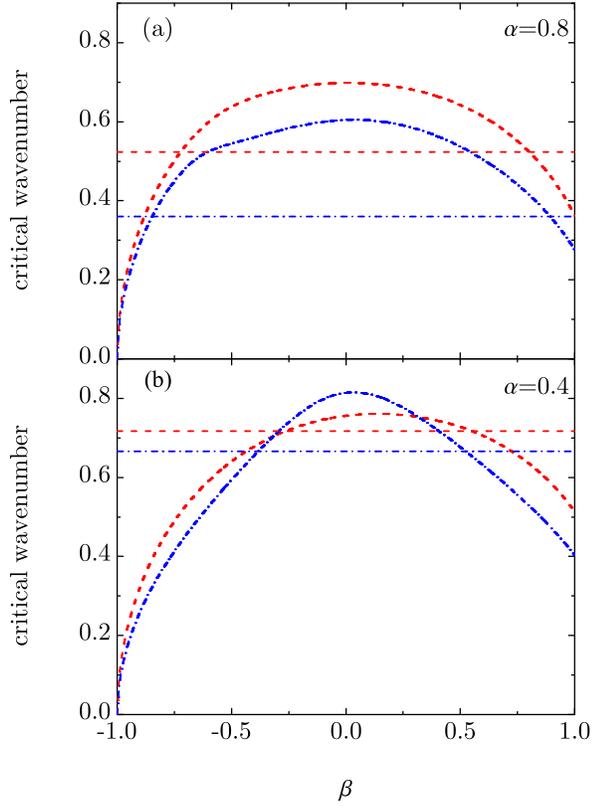}
\caption{Plot of the critical wave numbers $k_\perp$ (-- -- --) and $k_\|$ (--$\cdot$--$\cdot$--) as functions of $\beta$ for a reduced moment of inertia $\kappa=\frac{2}{5}$ with (a) $\alpha=0.8$ and (b) $\alpha=0.4$. The horizontal lines represent the respective values ($k_{\perp,\sm}$ and $k_{\|,\sm}$) in the  purely smooth case.}
\label{fig:kcritical}
\end{figure}

\begin{figure}[tbp]
\includegraphics[width=0.9\columnwidth]{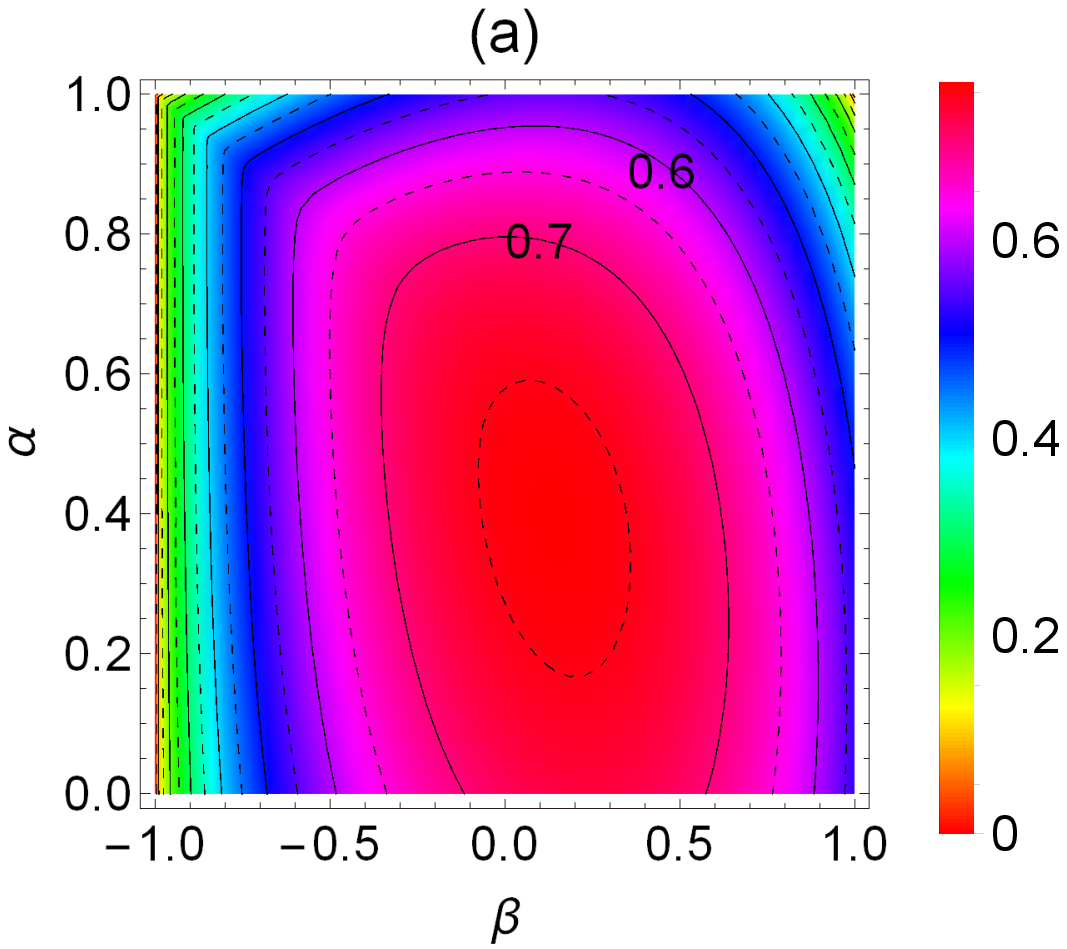}\\
\includegraphics[width=0.9\columnwidth]{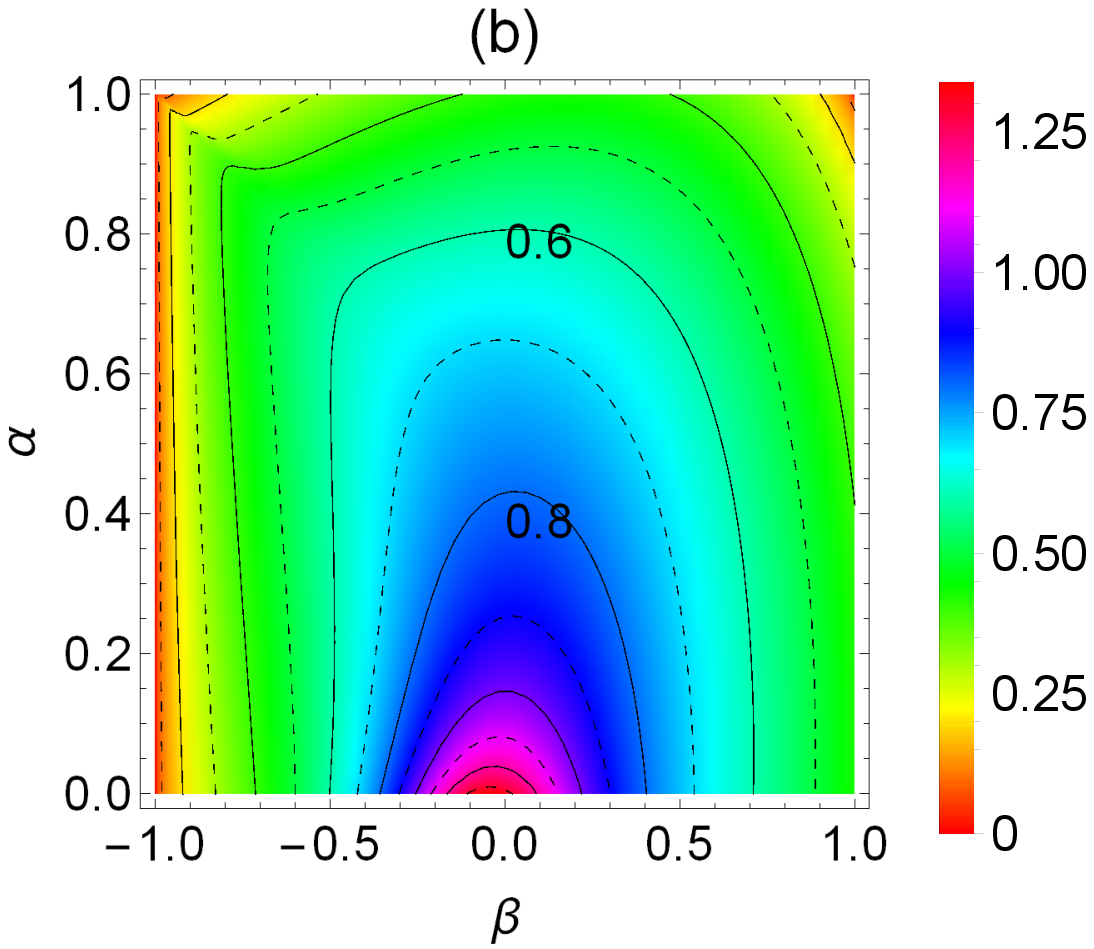}\\
\includegraphics[width=0.9\columnwidth]{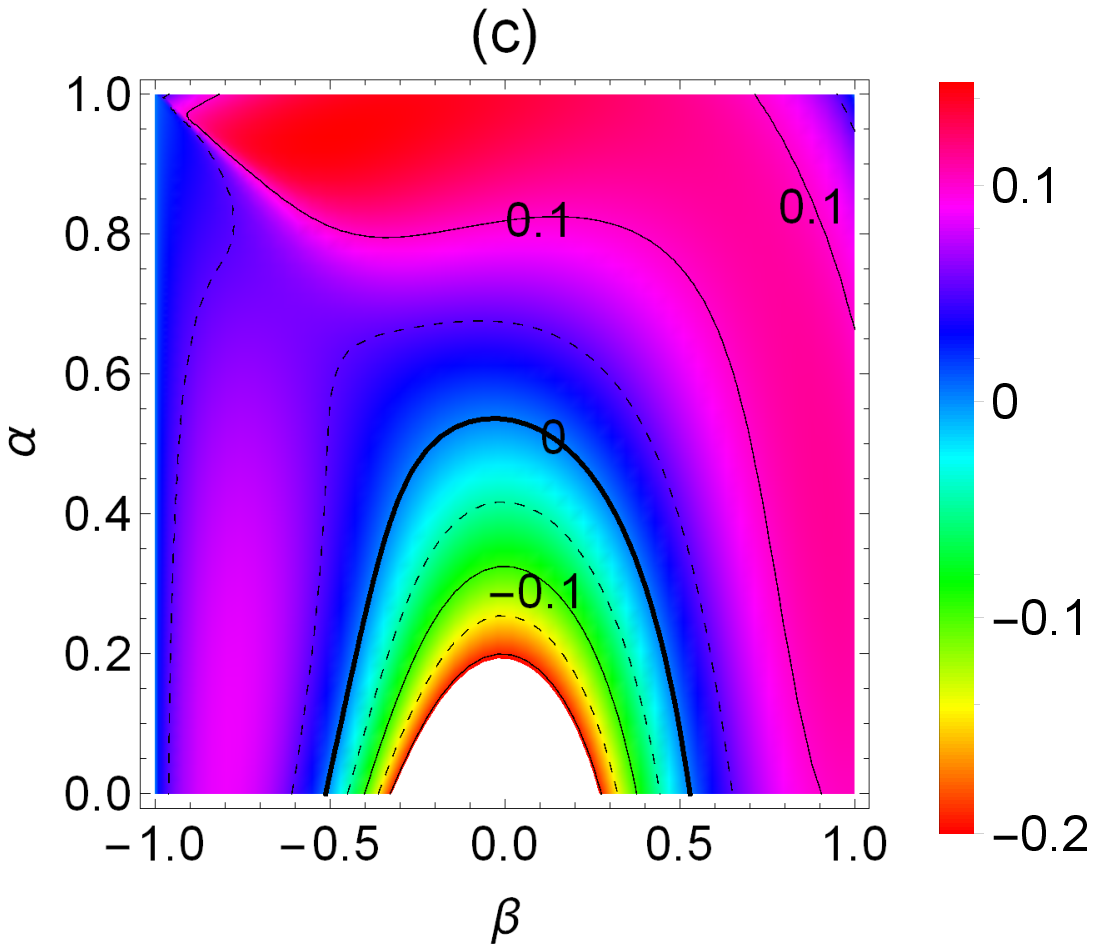}
\caption{Density plot of (a) the critical wave number $k_\perp$,  (b) the critical wave number $k_\|$, and (c) the difference $k_\perp-k_\|$, all in the case of uniform spheres ($\kappa=\frac{2}{5}$). Two adjacent contour lines correspond to a step (a) $\Delta k_\perp=0.05$, (b) $\Delta k_\|=0.1$, and (c) $\Delta (k_\perp-k_\|)=0.05$. In panel (c), the thick contour line $k_\perp-k_\|=0$ divides the plane $(\alpha,\beta)$  into the (outer) region where the  transversal shear mode is the most unstable one ($k_\perp>k_\|$) and the (inner) region where the longitudinal heat mode is the most unstable one ($k_\|>k_\perp$).}
\label{fig:kperp-kpar}
\end{figure}

\begin{figure}[tbp]
\includegraphics[width=0.9\columnwidth]{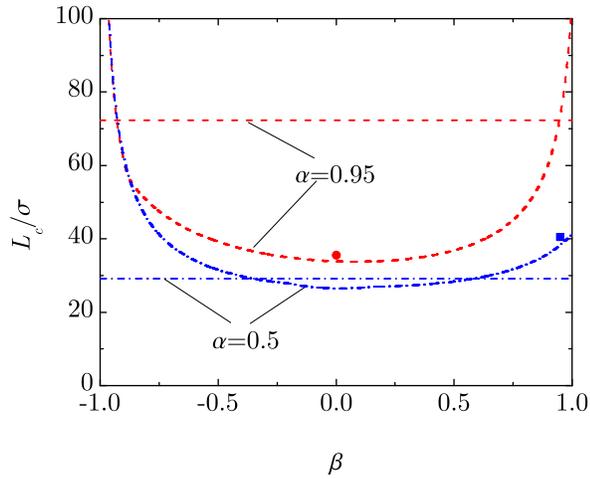}
\caption{Plot of the reduced critical length $L_c/\sigma$ as a function of $\beta$ for a reduced moment of inertia $\kappa=\frac{2}{5}$ and a solid volume fraction $\phi=0.05$ with $\alpha=0.95$ (-- -- --) and $\alpha=0.5$ (--$\cdot$--$\cdot$--) [see Eq.\ \eqref{3.25}]. The horizontal lines represent the respective values ($L_{c,\sm}/\sigma$) in the  purely smooth case, while the symbols are MD results at $(\alpha,\beta)=(0.95,0)$ (circle) and $(0.5, 0.95)$ (square) \cite{Mitrano}.}
\label{fig:Lcritical}
\end{figure}

\begin{figure}[tbp]
\includegraphics[width=0.9\columnwidth]{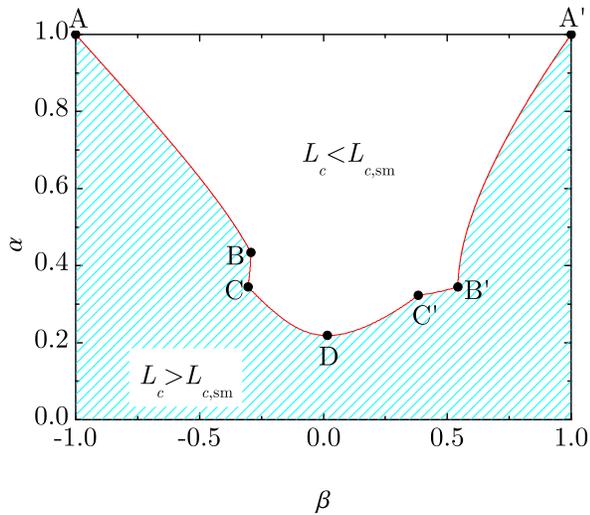}
\caption{Phase diagram in the case of uniform spheres ($\kappa=\frac{2}{5}$). The locus $L_c=L_{c,\sm}$ (solid line) splits the plane $(\alpha,\beta)$ into the hatched region, where  $L_c>L_{c,\sm}$ (so the instability is attenuated by friction), and the unhatched region,  where  $L_c<L_{c,\sm}$ (so the instability is enhanced by friction).}
\label{fig:atten}
\end{figure}

\begin{figure}[tbp]
\includegraphics[width=0.9\columnwidth]{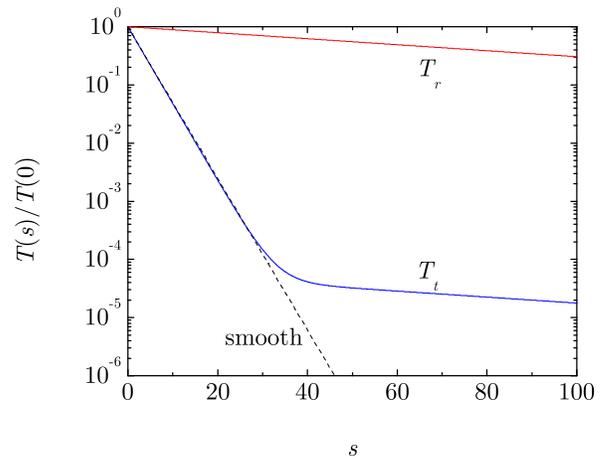}
\caption{Translational and rotational temperatures versus the scaled time variable $s$ [see Eq.\ \eqref{3.2}] for $\alpha=0.8$, $\kappa=\frac{2}{5}$, and $\beta=-0.99$. The initial condition is $T_r(0)=T_t(0)$. The dashed line represents the evolution of $T_t(s)$ in the purely smooth case ($\beta=-1$).}
\label{fig:quasi}
\end{figure}

\begin{figure}[tbp]
\includegraphics[width=0.9\columnwidth]{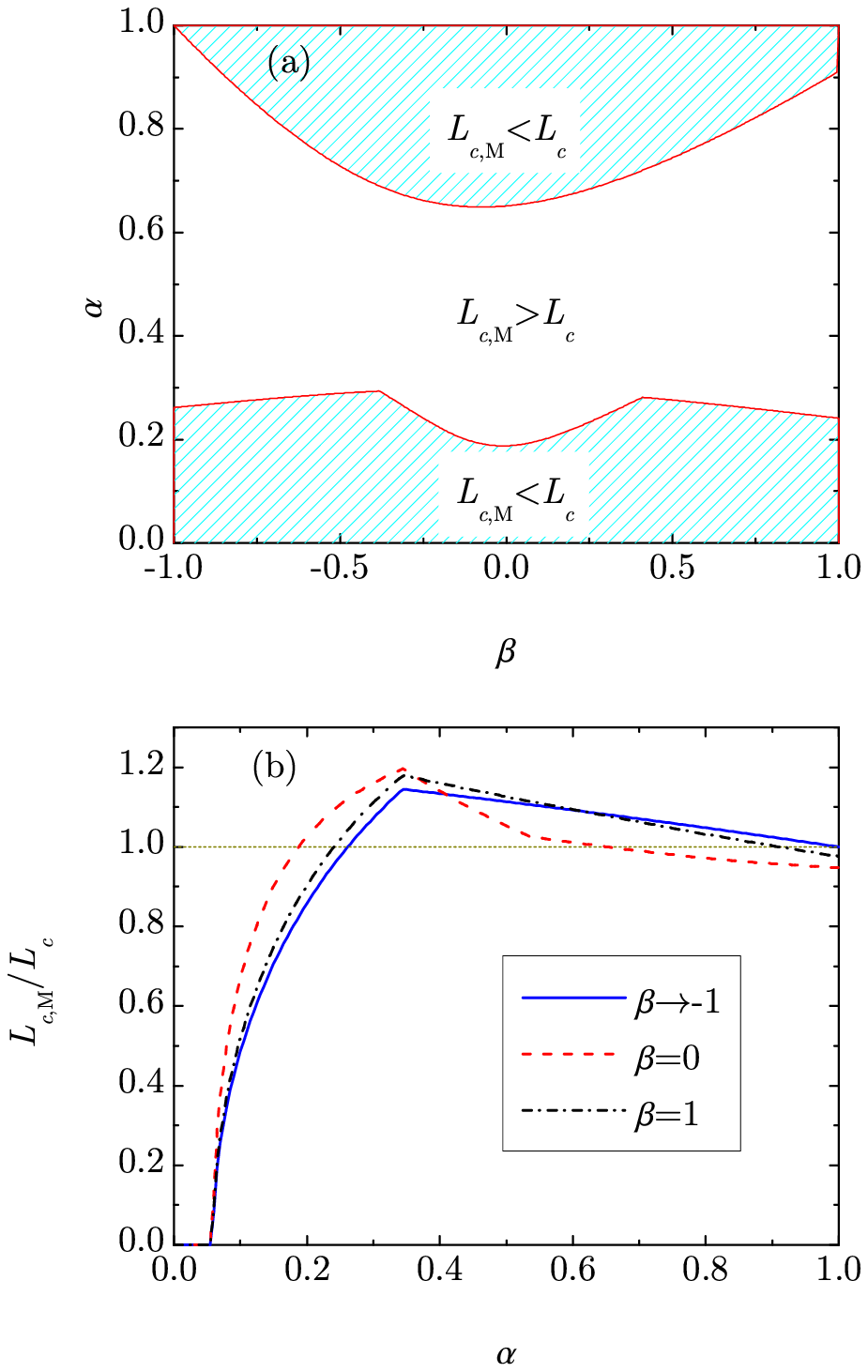}
\caption{(a) Phase diagram in the case of uniform spheres ($\kappa=\frac{2}{5}$) distinguishing the (hatched) upper and lower regions where $L_{c,\mitr}<L_c$ [see Eqs.\ \eqref{3.25} and \eqref{3.25mitr}] from the (unhatched) middle region where $L_{c,\mitr}>L_c$. (b) Plot of the ratio $L_{c,\mitr}/L_c$ versus $\alpha$ for $\beta\to -1$ (---), $\beta=0$ (-- -- --), and $\beta=1$ (--$\cdot$--$\cdot$--).}
\label{fig:comparison}
\end{figure}

\begin{figure}[tbp]
\includegraphics[width=0.9\columnwidth]{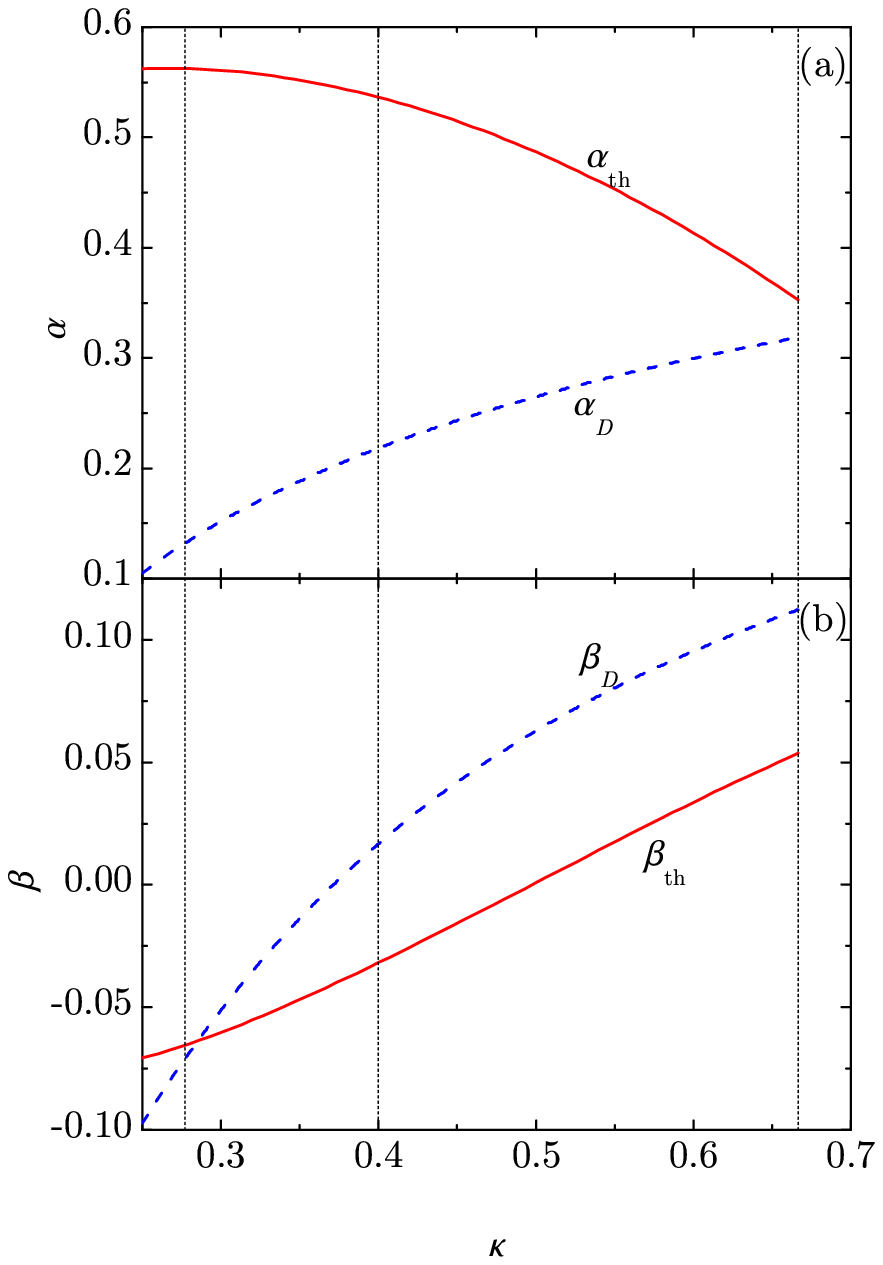}
\caption{Plot of (a) $\alpha_\thr$ and $\alpha_D$, and (b) $\beta_\thr$ and $\beta_D$ as functions of the reduced moment of inertia $\kappa$. The vertical dotted lines mark the values $\kappa=0.277$ (above which $k_\|$ is well defined for all $\alpha$ and $\beta$), $\kappa=\frac{2}{5}$ (uniform mass distribution), and $\kappa=\frac{2}{3}$ (mass concentrated on the spherical surface).}
\label{fig:alpha_th}
\end{figure}

\section{Discussion}
\label{sec4}

\subsection{Long-wavelength limit}
Before considering the general case ($k \neq 0$), it is convenient  to consider first the solutions to Eq.\ \eqref{3.14} in the long-wavelength limit $k\to 0$ (Euler hydrodynamic order). In this case, the eigenvalues are simply given by
\beq
\label{3.18}
\so_\perp(0)=\zeta^*, \quad \{\so_{||,a}(0);a=1,2,3\}=\left\{-\zeta^*,0,\zeta^*\right\}.
\eeq
Since two of the eigenvalues are positive (corresponding to growth of the initial perturbation in time), this means that some of the solutions are unstable, as expected. The zero eigenvalue represents a marginal stability solution, while the negative eigenvalue provides a stable solution. The unstable (positive) modes simply  emerge from the normalization of the flow velocity by the time-dependent thermal velocity $v_H(t)$, which is required to get time-independent coefficients in the linearized hydrodynamic equations after scaling the hydrodynamic fields with respect to their homogeneous values.

The solution to the cubic equation \eqref{3.14} for small wave numbers $k$ can be written in the form of the series expansion
\beq
\label{3.19}
\so_{\|,a}(k)=\so_{\|,a}^{(0)}+k^2 \so_{\|,a}^{(2)}+\cdots, \quad a=1,2,3,
\eeq
where the Euler eigenvalues $\so_{\|,a}^{(0)}=\so_{\|,a}(0)$ are given by the second equality in Eq.\ \eqref{3.18}. Note that only even powers in $k$ appear in Eq.\ \eqref{3.19}.  Substitution of the expansion \eqref{3.19} into Eq.\ \eqref{3.14} yields
\begin{subequations}
\label{3.19-22}
\beq
\label{3.21}
\so_{\|,1}^{(2)}=\frac{6-3\xi+\tau_t}{6\zeta^*}-\frac{5\lambda^*}{8},
\eeq
\beq
\label{3.20}
\so_{\|,2}^{(2)}=-\frac{1}{\zeta^*},
\eeq
\beq
\label{3.22}
\so_{\|,3}^{(2)}=-\frac{2\eta^*}{3}-\frac{\eta_b^*}{2}-\frac{\tau_t-3\xi}{6\zeta^*}.
\eeq
\end{subequations}
Notice that the small wave number solutions \eqref{3.19-22} are relevant since they are of the same order $\mathcal{O}(k^2)$ as the NS hydrodynamic equations.

\subsection{Finite wavelength}

By extending the usual nomenclature in normal fluids \cite{RL77} to the case of granular gases, the three longitudinal modes can be referred to as two ``sound'' modes ($\omega_{\|,1}$ and $\omega_{\|,2}$) and a ``heat'' mode ($\omega_{\|,3}$). As happens for smooth-sphere granular gases \cite{BDKS98,G05}, the shear and heat  modes ($\omega_\perp$ and $\omega_{\|,3}$) are real for all $k$. On the other hand, the two sound modes ($\omega_{\|,1}$ and $\omega_{\|,2}$) become a complex conjugate pair of propagating modes in a certain interval $k_*<k<k_{**}$, where $k_*$ and $k_{**}$ depend on $\alpha$, $\beta$, and $\kappa$. In particular, $k_{*}\to 0$ in the quasismooth limit $\beta\to -1$.

The dispersion relations $\so(k)$ are illustrated in Fig.\ \ref{fig:modes} for the case of uniform spheres ($\kappa=\frac{2}{5}$) at a representative value of the coefficient of normal restitution ($\alpha=0.8$). Figure \ref{fig:modes}(a) corresponds to the purely smooth case ($\beta=-1$) \cite{BDKS98,G05}, while Fig.\  \ref{fig:modes}(b) represents the quasismooth limit $\beta\to -1$. In Figs.\ \ref{fig:modes}(c)--\ref{fig:modes}(f) roughness increases from $\beta=-0.5$ to $\beta=1$. The respective values of $(k_{*},k_{**})$ are (a) $(0.07,1.71)$, (b) $(0,0.69)$, (c) $(0.12,2.11)$, (d) $(0.15,10.54)$, (e) $(0.12,8.54)$, and (f) $(0.03,26.72)$. Apart from the instability associated with the transversal shear mode, Eq.\ \eqref{3.11}, Fig.\ \ref{fig:modes} highlights that the heat mode also becomes unstable for $k<k_\parallel$, where
\beq
\label{3.23}
k_\parallel=\sqrt{\frac{8\zeta^*}{5\left(\lambda^*-\mu^*\right)}}
\eeq
can be obtained from Eq.\ \eqref{3.14} by setting $\so=0$. The only exception is the quasismooth limit ($\beta\to -1$) since in that case $\zeta^*\to 0$ and hence both $k_\perp$ and $k_\|$ tend to zero.

For purely smooth spheres ($\beta=-1$), the critical wave number $k_{\|,\sm}$ can be obtained from Eq.\ \eqref{3.23} with the replacements $\zeta^*=\zeta^*_\sm$, $\lambda^*=2\lambda^*_\sm$, and $\mu^*=2\mu^*_\sm$. In that case, one has $k_{\perp,\sm}>k_{\|,\sm}$ if $\alpha$ is higher than a threshold value $\alpha_{\thr,\sm}=0.345$ \cite{G05,MGHEH12}, as illustrated in Fig.\ \ref{fig:modes}(a) for $\alpha=0.8$. Thus, in the purely smooth case, the instability is driven by the transversal shear mode (vortex instability) in the inelasticity domain $\alpha>\alpha_{\thr,\sm}$. Figure \ref{fig:modes} shows that this is also the case for rough spheres with $\alpha=0.8$.
The property $k_\perp>k_\|$ for $\alpha=0.8$ is clearly seen in Fig.\ \ref{fig:kcritical}(a), which displays $k_\perp$ and $k_\|$ as functions of $\beta$.
On the other hand, if $\alpha$ is small enough, there exists an intermediate window of $\beta$-values where $k_\|>k_\perp$, so the instability is driven in that case by the longitudinal heat mode (clustering instability). This is the case shown in Fig.\ \ref{fig:kcritical}(b) for $\alpha=0.4$. If the spheres are uniform  ($\kappa=\frac{2}{5}$), the threshold value of $\alpha$ below which it is possible to have $k_\|>k_\perp$ is $\alpha_\thr=0.536$. This threshold value is about $55\%$ times larger than the one ($\alpha_{\thr,\sm}=0.345$) for purely smooth spheres.

Apart from $k_\perp$ and $k_\|$ for rough spheres, Fig.\ \ref{fig:kcritical} also includes as horizontal lines those quantities ($k_{\perp,\sm}$ and $k_{\|,\sm}$) for purely smooth spheres. As can be seen, one has $k_\perp<k_{\perp,\sm}$ and $k_\|<k_{\|,\sm}$ in the regions of small and large roughness, while the opposite happens for intermediate roughness.
This explains the \emph{dual role} of roughness previously observed in MD simulations \cite{MDHEH13}: ``high levels of friction actually attenuate instabilities relative to the frictionless case, whereas moderate levels enhance instabilities compared to frictionless systems.''
However, as will be seen below, the attenuation effect dominates for any level of friction if the inelasticity is high enough.

The full combined dependence of $k_\perp$, $k_\|$, and $k_\perp-k_\|$ on both $\alpha$ and $\beta$ is illustrated in Fig.\ \ref{fig:kperp-kpar}, again for $\kappa=\frac{2}{5}$. Figures \ref{fig:kperp-kpar}(a) and \ref{fig:kperp-kpar}(b) confirm that, in agreement with Fig.\ \ref{fig:kcritical}, the critical wave numbers $k_\perp$ and $k_\|$ reach their maximum values at intermediate roughness ($\beta\approx 0$) for a given value of $\alpha$. On the other hand, while at $\beta\approx 0$ the longitudinal critical wave number $k_\|$ monotonically increases with increasing inelasticity, the transversal critical wave number $k_\perp$ presents a nonmonotonic dependence on $\alpha$ at $\beta\approx 0$. As a consequence, and as already anticipated in Fig.\ \ref{fig:kcritical}, the vortex instability (associated with $k_\perp$) is preempted by the clustering instability (associated with $k_\|$) in a dome-shaped region with an apex at $(\alpha,\beta)=(\alpha_\thr,\beta_\thr)=(0.536,-0.032)$ [see the thick contour line in Fig.\ \ref{fig:kperp-kpar}(c)].

\subsection{Critical length}
In a system with periodic boundary conditions, the smallest allowed wave number is $2\pi/L$, where $L$ is the largest system length. Thus, we can identify a critical length $L_{c}$, such that the system becomes unstable when $L>L_\text{c}$. According to the scaling \eqref{3.2}, the value of $L_{c}$ is given by
\begin{equation}
\label{3.24}
{L_c}=\frac{4\pi v_H}{\nu_H}\min\{k_{\perp}^{-1},k_\parallel^{-1}\}.
\end{equation}
Making use of Eqs.\ \eqref{2.10} and \eqref{v_H}, this can be rewritten as
\beq
\label{3.25}
\frac{L_{c}}{\sigma}=\frac{5\pi\sqrt{\pi}}{24\phi\chi(\phi)}\min\{k_{\perp}^{-1},k_\parallel^{-1}\} .
\eeq

Figure \ref{fig:Lcritical} shows the $\beta$-dependence of $L_c/\sigma$ for $\alpha=0.95$ and $\alpha=0.5$ (both with $\kappa=\frac{2}{5}$) at a small (but nonzero) solid volume fraction $\phi=0.05$, in which case $\chi(\phi)=1.137$. The respective values for purely smooth spheres, $L_{c,\sm}/\sigma$, are represented by horizontal lines. The dual role of roughness (or friction) is again quite apparent. For small and large roughness the system is less unstable (instability attenuation) than its smooth-sphere counterpart, while the opposite happens (instability enhancement) for intermediate roughness. It can be observed that the central enhancement  region is much shorter with $\alpha=0.5$ than for $\alpha=0.95$. Figure \ref{fig:Lcritical} also includes values obtained by MD simulations \cite{Mitrano} for a representative case of instability enhancement ($\alpha=0.95$, $\beta=0$) and a representative case of instability attenuation ($\alpha=0.5$, $\beta=0.95$). As can be observed, the agreement with the theoretical predictions is excellent in both cases.

As said above, Fig.\ \ref{fig:Lcritical} shows that the medium-roughness region where $L_c<L_{c,\sm}$ shrinks as inelasticity increases (i.e., as $\alpha$ decreases). Is a threshold value of $\alpha$ reached below which $L_c>L_{c,\sm}$ for all $\beta$? To address this question, Fig.\ \ref{fig:atten} displays a phase diagram (assuming uniform spheres, i.e., $\kappa=\frac{2}{5}$) where the enhancement region (or ``phase'') $L_c<L_{c,\sm}$ is separated from the attenuation region (or ``phase'') $L_c>L_{c,\sm}$ by the locus line $L_c=L_{c,\sm}$. The peculiar shape of the locus is due to the change from the shear mode to the heat mode as the most unstable one as inelasticity increases. More specifically, in the segments A--B and A'--B' one has $(L_c,L_{c,\sm})\propto (k_\perp^{-1},k_{\perp,\sm}^{-1})$, while in the segments B--C and B'--C' the situation is $(L_c,L_{c,\sm})\propto (k_\|^{-1},k_{\perp,\sm}^{-1})$ and $(L_c,L_{c,\sm})\propto (k_\perp^{-1},k_{\|,\sm}^{-1})$, respectively. Finally, in the segment C--D--C',  $(L_c,L_{c,\sm})\propto (k_\|^{-1},k_{\|,\sm}^{-1})$. The coordinates of the relevant points are $(\alpha_{B},\beta_{B})=(0.434,-0.293)$, $(\alpha_{B'},\beta_{B'})=(0.345,0.543)$, $(\alpha_{C},\beta_{C})=(0.345,-0.305)$, $(\alpha_{C'},\beta_{C'})=(0.323,0.383)$, and $(\alpha_{D},\beta_{D})=(0.218,0.017)$.
Note that $\alpha_B<\alpha_{\thr}$, while $\alpha_{B'}=\alpha_{C}=\alpha_{\thr,\sm}$.
As Fig.\ \ref{fig:atten} clearly shows, the locus $L_c=L_{c,\sm}$ presents a minimum vertex at point D. Therefore, if $\alpha<\alpha_D$, the frictional system is always less unstable than the frictionless system.

\subsection{On the quasismooth limit}
\label{sec4.D}

It seems paradoxical that, as illustrated in Figs.\ \ref{fig:modes}(a), \ref{fig:modes}(b), \ref{fig:kcritical}, \ref{fig:Lcritical}, and \ref{fig:atten}, the HCS is stable in the limit $\beta\to -1$, while it might be unstable (if $L>L_{c,\sm}$) in the purely smooth case. The explanation lies in the fact that the HCS with $\beta=-1$ is very different from the one with $\beta\gtrsim -1$. In the former case, the rotational degrees of freedom do not play any role at all and the translational temperature decays with the cooling rate \eqref{zeta_smooth}. On the other hand, in the quasismooth case the rotational and translational degrees of freedom are coupled,  the temperature ratio is approximately given by Eq.\ \eqref{theta_quasi}, and both temperatures decay with a much smaller cooling rate given by Eq.\ \eqref{zeta_quasi}.

The interesting question is whether the transient regime \emph{prior} to the HCS for $\beta\gtrsim -1$ might actually be unstable. As shown by Luding \emph{et al.} \cite{LHMZ98}, the initial decay of the translational
temperature $T_t$ of nearly smooth particles is dominated by the coefficient of normal restitution $\alpha$ and reaches a very small value (relative to $T_r$) before the asymptotic HCS is attained. This is illustrated by Fig.\ \ref{fig:quasi}, which shows the evolution of $T_t$ and $T_r$ for $\alpha=0.8$, $\kappa=\frac{2}{5}$, and $\beta=-0.99$,
as obtained by numerically solving the coupled set of equations $\dot{T}_{t}=-\zeta_{t}T_{t}$ and $\dot{T}_{r}=-\zeta_{r}T_{r}$,
starting from an initial condition of energy equipartition, i.e., $T_r(0)=T_t(0)$. Here, $\zeta_t$ and $\zeta_r$, with $\zeta=(\zeta_t T_t+\zeta_r T_r)/(T_t+T_r)$, are the collisional rates of change of $T_t$ and $T_r$, respectively \cite{KSG14}.
It can be observed that up to $s\approx 30$ collisions per particle the evolution of $T_t$ is practically indistinguishable from that of the smooth-sphere system. Therefore, a perturbation with a wavelength larger than $L_{c,\sm}$ during this transient regime will develop vortex or cluster instabilities.
In order to prevent instabilities in the quasismooth limit, one would need to fine-tune the initial state with a temperature ratio $T_r(0)/T_t(0)\sim \theta$.

\subsection{Comparison with the approach of Mitrano \emph{et al.} \protect\cite{MDHEH13}}

According  to the heuristic approach of Mitrano \emph{et al.}  \cite{MDHEH13}, the critical wave numbers are obtained by starting from the smooth-sphere values $k_{\perp,\sm}$ and $k_{\|,\sm}$, and simply replacing the smooth-sphere cooling rate $\zeta^*_\sm$ [see Eq.\ \eqref{zeta_smooth}] by the rough-sphere cooling rate $\zeta^*$ [see Eq.\ \eqref{2.13}]:
\beq
\label{3.11mitr}
k_{\perp,\mitr}=\sqrt{\frac{2\zeta^*}{\eta^*_\sm}},\quad
k_{\parallel,\mitr}=\sqrt{\frac{4\zeta^*}{5\left(\lambda^*_\sm-\mu^*_\sm\right)}}.
\eeq
The associated critical length is then
\beq
\label{3.25mitr}
\frac{L_{c,\mitr}}{\sigma}=\frac{5\pi\sqrt{\pi}}{24\phi\chi(\phi)}\min\{k_{\perp,\mitr}^{-1},k_{\parallel,\mitr}^{-1}\} .
\eeq

Figure \ref{fig:comparison}(a) shows that, at a given value of $\beta$, $L_{c,\mitr}<L_c$ if $\alpha$ is either high enough or low enough. In those cases, the simple proposal \eqref{3.11mitr} predicts the system to be more unstable than the more sophisticated analysis based on the true transport coefficients. The opposite happens in a window of intermediate values of $\alpha$. On the other hand, a quantitative comparison shows that $L_{c,\mitr}\approx L_c$ in the upper region ($L_{c,\mitr}<L_c$) and in most of the middle region ($L_{c,\mitr}<L_c$), while the limitations of the simple approach \eqref{3.25mitr} show up around $\alpha\approx \alpha_{\thr,\sm}=0.345$ and in the lower region ($L_{c,\mitr}<L_c$).  This is illustrated in Fig.\ \ref{fig:comparison}(b) for the quasismooth limit $\beta\to -1$, the medium roughness case $\beta=0$, and the completely rough case $\beta=1$.

\subsection{Influence of the moment of inertia}
All the results displayed in Figs.\ \ref{fig:modes}--\ref{fig:comparison} correspond to spheres with a uniform mass distribution ($\kappa=\frac{2}{5}$). Nevertheless, at least at a quantitative level, the results are expected to be influenced by the value of the reduced moment of inertia $\kappa$.
In this respect, it must be mentioned that the critical longitudinal wave number \eqref{3.23} turns out to be well defined for all $\alpha$ and $\beta$ only if $\kappa>0.277$. For smaller values of the reduced moment of inertia, however, $\lambda^*<\mu^*$ below a certain $\kappa$-dependent value of $\alpha$ (with a maximum $\alpha=0.2533$ at $\kappa\to 0$) and for a certain interval of values of $\beta$. A similar situation takes place in the purely smooth case, where $\lambda_\sm^*<\mu_\sm^*$ if $\alpha<0.0588$. This is likely an artifact due to the limitations of the Sonine approximation for extremely low values of $\alpha$ and/or $\kappa$.

In order to assess the influence of $\kappa$, the threshold values $\alpha_\thr$ (below which one may have $k_\|>k_\perp$) and $\alpha_D$ (below which $L_c>L_{c,\sm}$ for all $\beta$) are plotted as functions of $\kappa\geq 0.25$ in Fig.\ \ref{fig:alpha_th}(a). Figure \ref{fig:alpha_th}(b) does the same but for the other coordinates $\beta_\thr$ and $\beta_D$. As the mass of the spheres becomes more concentrated in the inner layers (i.e., as $\kappa$ decreases), $\alpha_\thr$ tends to increase  (although it presents a maximum value $\alpha_\thr=0.5628$ at $\kappa=0.2656$), while $\alpha_D$ monotonically decreases. As for the associated values $\beta_\thr$ and $\beta_D$ of the coefficient of tangential restitution, both decrease as $\kappa$ decreases (especially in the case of $\beta_D$), but their magnitudes are smaller than about $0.1$, so they always lie in the region of medium roughness $\beta\approx 0$.

\section{Conclusions}
\label{sec5}

In this paper we have undertaken a rather complete study of the linear stability conditions of the HCS of a dilute granular gas modeled as a system of (identical) inelastic and frictional hard spheres. Inelasticity  and surface friction are characterized by constant coefficients of normal ($\alpha$) and tangential ($\beta$) restitution. The analysis is based on the NS hydrodynamic equations, linearized around the time-dependent HCS solution. The most relevant outcome is the determination of the critical length $L_c$, such that the system is linearly unstable for sizes larger than $L_c$. The novel aspect of our study, not accounted for in previous ones \cite{MDHEH13}, is the use of the detailed nonlinear dependence of the NS transport coefficients on both $\alpha$ and $\beta$, as well as on the reduced moment of inertia $\kappa$ \cite{KSG14}. This allows one to explore the impact of roughness on the critical length $L_c$ without \emph{a priori} any restriction on $\alpha$, $\beta$, and $\kappa$.

As in the purely smooth case \cite{BDKS98,G05}, two of the five hydrodynamic modes (the two longitudinal ``sound'' modes) are stable. On the other hand, a doubly degenerate transversal (shear) mode and a longitudinal ``heat'' mode become unstable for (reduced) wave numbers smaller than certain critical values $k_\perp$ and $k_\|$, respectively. In general, the instability is driven by the transversal mode, i.e., $k_\perp>k_\|$. On the other hand, if $\alpha$ is small enough ($\alpha<\alpha_\thr=0.536$ in the case of uniform spheres) and $\beta$ lies inside an $\alpha$-dependent interval around $\beta\approx 0$, the situation is reversed, i.e., the heat mode is the most unstable one. As a consequence, the critical length $L_c\propto \min\{k_\perp^{-1},k_\|^{-1}\}$ exhibits a nontrivial dependence on $\alpha$, $\beta$, and $\kappa$. Comparison of the theoretical predictions for $L_c$ against preliminary MD simulations \cite{Mitrano} shows an excellent agreement.

An interesting point is the comparison between the critical length $L_c$ of a gas of rough spheres and its corresponding counterpart, $L_{c,\sm}$, of a gas of purely smooth spheres. One could naively expect that the existence of friction would enhance the instability of the HCS, namely $L_c<L_{c,\sm}$, for a common value of the inelasticity parameter $\alpha$. Nevertheless, this is not the general case. As illustrated in Fig.\ \ref{fig:atten} for uniform spheres, an attenuation effect is present at sufficiently low or sufficiently high levels of friction. This dual role of friction was already observed by Mitrano \emph{et al.} \cite{MDHEH13} in MD simulations and in a simple kinetic theory description for dense gases. On the other hand, the enhancement middle region of roughness  disappears if the inelasticity is large enough ($\alpha<\alpha_D=0.218$ for uniform spheres).
With respect to the influence of the moment of inertia, our results show that, as the mass of each sphere concentrates nearer the surface (i.e., as $\kappa$ increases), the values of $\alpha_\thr$ and $\alpha_D$  decrease and increase, respectively.

It is worthwhile mentioning that the cooling rate of the gas of rough spheres, as compared with that of purely smooth spheres, already exhibits a dual behavior: dissipation of energy is enhanced by friction only if $\alpha$ is larger than a certain threshold value ($\alpha>0.401$ for uniform spheres) and $\beta$ lies in an $\alpha$-dependent interval around $\beta\approx 0$. Otherwise, friction attenuates energy dissipation. This explains the good performance of the simple kinetic theory proposed by Mitrano \emph{et al.} \cite{MDHEH13}, where the critical wave numbers $k_\perp$ and $k_\|$ are obtained by assuming the same expressions as for the purely smooth system, except for the replacement of the cooling rate.

A subtle point is the suppression of clustering and vortex formation in the quasismooth limit ($\beta\to -1$). As explained in Sec.\ \ref{sec4.D}, if the initial state is not close enough to the HCS of rough spheres, the transient regime can be almost indistinguishable from the HCS of the purely smooth-sphere gas and thus instabilities might occur before the asymptotic HCS is reached.

Additionally, it must be noted that all the results obtained in this paper have been obtained in the context of a collision
model where both coefficients of restitution are independent of the impact velocity.
Conversely, results derived with a viscoelastic model where $\alpha$ tends to $1$ as the impact velocity decreases show that structure formation occurs in
free granular gases only as a transient phenomenon, whose duration increases with the system size \cite{BSSP04}.
However, the experimental measurement of $\alpha$ at very small impact velocities is very challenging \cite{DF17}. Some independent experiments \cite{SLCL09,GBG09} provide evidence on a sharp decrease of $\alpha$ at small impact velocities, possibly due to van der Waals attraction at relatively low surface energies for typical grain materials.

Finally, we hope that the results presented in this work will stimulate the performance of computer simulations to further assess their practical usefulness.

\begin{acknowledgments}
We want to thank Peter P.\ Mitrano for making the simulation data included in Fig.\ \ref{fig:Lcritical} available to us. V.G. and A.S. acknowledge the financial support of the Ministerio de Econom\'ia y Competitividad (Spain) through Grant No.\ FIS2016-76359-P, partially financed by ``Fondo Europeo de Desarrollo
Regional'' funds. The research of G.M.K is supported by Conselho Nacional de Desenvolvimento Cient\'ifico
e Tecnol\'ogico (CNPq), Brazil, through Grant No. 303251/2015-8.
\end{acknowledgments}

\appendix*
\section{Explicit expressions for the NS transport and cooling-rate  coefficients}
\label{appA}

In this Appendix, the expressions for the NS transport coefficients ($\eta^*$, $\eta_b^*$, $\lambda^*$, $\mu^*$) and the NS cooling rate coefficient ($\xi$) are explicitly given.

The reduced coefficients $\eta^*$ and $\eta_b^*$ associated with the pressure tensor are given by Eq.\ \eqref{2.14} with
\begin{subequations}
\label{a1-a2}
\beq
\label{a1}
\nu_\eta^*=(\at+\bt)(2-\at-\bt)+\frac{\bt^2\theta}{6\kappa},
\eeq
\beq
\label{a2}
\gamma_E=\frac{2}{3}\frac{1}{\Xi_t-\Xi_r-\zeta^*},
\eeq
\end{subequations}
where
\begin{subequations}
\label{a5-a4}
\beq
\label{a5}
\at=\frac{1+\al}{2}, \quad \bt=\frac{1+\beta}{2}\frac{\kappa}{1+\kappa},
\eeq
\beq
\label{a3}
\Xi_t=\frac{5}{8}\tau_r\Big[1-\al^2+(1-\beta^2)\frac{\kappa}{1+\kappa}-\frac{\kappa}{3}({\theta-5})
\left(\frac{1+\beta}{1+\kappa}\right)^2
\Big],
\eeq
\beq
\label{a4}
\Xi_r=\frac{5}{8}\tau_t\frac{1+\beta}{1+\kappa}\left[\frac{\theta-2}{3}(1-\beta)
+\frac{\kappa}{3}({\theta-5})\frac{1+\beta}{1+\kappa}\right].
\eeq
\end{subequations}

The reduced coefficients $\lambda^*$ and $\mu^*$ associated with the heat flux are given by Eq.\ \eqref{2.15} with
\begin{subequations}
\label{a6-a9}
\beq
\label{a6}
\gamma_{A_t}=\frac{Z_r-Z_t-2\zeta^*}{\left(Y_t-2\zeta^*\right)\left(Z_r-2\zeta^*\right)-Y_r Z_t},
\eeq
\beq
\label{a7}
\gamma_{A_r}=\frac{Y_t-Y_r-2\zeta^*}{\left(Y_t-2\zeta^*\right)\left(Z_r-2\zeta^*\right)-Y_r Z_t},
\eeq
\beq
\label{a8}
\gamma_{B_t}=\zeta^*\frac{\gamma_{A_t}\left(Z_r-\frac{3}{2}\zeta^*\right)-\gamma_{A_r}Z_t}
{\left(Y_t-\frac{3}{2}\zeta^*\right)\left(Z_r-\frac{3}{2}\zeta^*\right)-Y_r Z_t},
\eeq
\beq
\label{a9}
\gamma_{B_r}=\zeta^*\frac{\gamma_{A_r}\left(Y_t-\frac{3}{2}\zeta^*\right)-\gamma_{A_t}Y_r}
{\left(Y_t-\frac{3}{2}\zeta^*\right)\left(Z_r-\frac{3}{2}\zeta^*\right)-Y_r Z_t}.
\eeq
\end{subequations}
In Eqs.\ \eqref{a6-a9}, we have introduced the quantities
\begin{subequations}
\label{a10-a13}
\beq
\label{a10}
Y_t=\frac{41}{12}\left(\at+\bt\right)-\frac{33}{12}\left(\at^2+\bt^2\right)-\frac{4}{3}\at\bt-\frac{7}{12}
\frac{\theta \bt^2}{\kappa},
\eeq
\beq
\label{a12}
Y_r=\frac{25}{36}\frac{\bt}{\kappa}\left(1-3\frac{\bt}{\theta}-\frac{\bt}{\kappa}\right),
\eeq
\beq
\label{a11}
Z_t=-\frac{5}{6}\frac{\theta\bt^2}{\kappa},
\eeq
\beq
\label{a13}
Z_r=\frac{5}{6}\left(\at+\bt\right)+\frac{5}{18}\frac{\bt}{\kappa}\left(7-3\frac{\bt}{\kappa}-6{\bt}-
 4{\at}\right).
 \eeq
 \end{subequations}

Finally, the first-order contribution $\xi$ to the cooling rate is given by Eq.\ \eqref{2.13.1}, where $\gamma_E$ is given by Eq.\ \eqref{a2} and $\Xi$ is
\beq
\label{a14}
\Xi=\frac{5}{16}\tau_t\tau_r\left[1-\al^2+(1-\beta^2)\left(1+\frac{1}{3}\frac{\theta-5}{1+\kappa}\right)\right].
\eeq

\bibliography{D:/Dropbox/Mis_Dropcumentos/bib_files/Granular}
\end{document}